\documentclass[reprint,aps,prl,showpacs,amsmath,amssymb]{revtex4-1}
\usepackage{graphicx}
\usepackage{amsmath}
\usepackage{tabularx}
\usepackage{xfrac}
\usepackage{color}
\usepackage{bm}
\newcolumntype{Y}{>{\centering\arraybackslash}X}

\def\rhobo{\mbox{\boldmath{$\rho$}}}

\begin{document}
\title{S-wave elastic scattering of {\it o}-Ps from $\text{H}_2$ at low energy}
\author{J.-Y. Zhang$^{1-3,\ast}$\email{jzhang@wipm.ac.cn}, M.-S. Wu$^{1}$, Y. Qian$^{4}$, X. Gao$^{2}$, Y.-J. Yang$^{5}$,
  K. Varga$^{6}$, Z.-C. Yan$^{1,7}$,  and U. Schwingenschl\"{o}gl$^3$}
\affiliation{$^1$ State Key Laboratory of Magnetic Resonance and Atomic and Molecular Physics, Wuhan Institute of Physics and Mathematics, Chinese Academy of Sciences, Wuhan 430071, China}
\affiliation{$^2$ Beijing Computational Science Research Center, Beijing 100193, China}
\affiliation{$^3$ Physical Science and Engineering Division (PSE),
King Abdullah University of Science and Technology (KAUST), Thuwal
23955-6900, Saudi Arabia}
\affiliation{$^4$\,Department of Computer Science and Technology,
East China Normal University, Shanghai 200062, China}
\affiliation{$^5$\,Institute of Atomic and Molecular Physics,
Jilin University, Changchun 130012, China}
\affiliation{$^6$ Department of Physics and Astronomy, Vanderbilt University, Nashville, Tennessee 37235, USA}
\affiliation{$^7$ Department of Physics, University of New
Brunswick, Fredericton, New Brunswick, Canada E3B 5A3 }
\date{\today}
\begin{abstract}
The confined variational method is applied to investigate the low-energy elastic scattering
of ortho-positronium from $\text{H}_2$ by first-principles quantum mechanics.
Describing the correlation effect with explicitly correlated Gaussians, we obtain
accurate $S$-wave phase shifts and pick-off annihilation parameters for different
incident momenta. By a least-squares fit of the data to the effective-range theory, we
determine the $S$-wave scattering length, $A_s=2.06a_0$, and the zero-energy value of
the pick-off annihilation parameter, $^1\!\text{Z}_\text{eff}=0.1858$. The obtained
$^1\!\text{Z}_\text{eff}$ agrees well with the precise experimental value of $0.186(1)$
(J.\ Phys.\ B \textbf{16}, 4065 (1983)) and the obtained $A_s$ agrees well with the
value of $2.1(2)a_0$ estimated from the average experimental momentum-transfer cross
section for Ps energy below 0.3 eV (J.\ Phys.\ B \textbf{36}, 4191 (2003)).
\end{abstract}


\maketitle

{\it Introduction}. Scattering of Positronium (Ps), {\it i.\ e.}, a hydrogen-like atom composed of
electron and positron, from atoms and molecules is fundamentally important for
understanding the interaction between matter and antimatter
\cite{mcnutt79a,wright83a,zafar96a,skalsey98a,blackwood99a,biswas00b,armitage02a,nagashima03a,skalsey03a,
engbrecht08a,brawley10a,walters10a,wada10a,fabrikant14a,woods15a,wilde15a,fabrikant17a}.
Ps can be in a spin singlet state ({\it para}-positronium; {\it p}-Ps) or a spin triplet
state ({\it ortho}-positronium; {\it o}-Ps). Pick-off quenching is the process that the
positron in the {\it o}-Ps annihilates on collision with a molecular electron in the
opposite spin state. The accuracy of experimental determination of the pick-off
annihilation parameter $^1\!Z_{\rm eff}$ of {\it o}-Ps interaction with different
targets such as H$_2$, CH$_4$, and CO$_2$ \cite{mcnutt79a,wright83a,wada10a} is far
ahead of that achieved by theoretical methods. The experimental results therefore can be
used to test the quality and efficiency of theoretical methods, in particular the
accuracy of the generated scattering wave functions. The complicated short-range
electron-positron and electron-electron correlations as well as the electron exchange
between Ps and target play key roles in the low-energy scattering of Ps. Theoretically,
however, the accurate description of these interactions is very difficult and tedious
due to the complex nature of a multi-centre scattering system.

In this Letter, we present confined variational studies of the low energy scattering
properties of the experimentally studied {\it o}-Ps-H$_2$ system. The work extends the
{\it ab-initio} theoretical description of the scattering of a composite projectile from
a one-center target to a multi-center target. The obtained zero-energy value of the
pick-off annihilation parameter, which is calculated for the first time ever, and the
scattering length show excellent agreement with experiments \cite{wright83a,nagashima03a},
demonstrating the high accuracy of the confined variational method (CVM).

The CVM \cite{mitroy08d,zhang09b,zhang12c} was first developed by Mitroy {\it et al.}
to accurately determine phase shifts of the low-energy elastic scattering of electrons
$(e^{-})$ or positrons ($e^{+}$) from few-$e^{-}$ atoms in 2008. In 2012, the CVM was
further developed by Zhang {\it et al.} \cite{zhang12c} to study the scattering of
projectiles with internal structure, such as Ps. The CVM phase shifts for the S-wave
$e^-$-He scattering at wave number $k=0.2a^{-1}_0$ and for the S-wave Ps-H elastic scattering at
$k=0.1a^{-1}_0$ and $k=0.2a^{-1}_0$ have set a benchmark for other theoretical methods~\cite{woods15a}.
In addition, the CVM was used to generate basis sets of energy-optimized explicitly
correlated Gaussian (ECG) functions for other collision calculation methods such as the stabilization
method \cite{zhang09b} and Kohn variational method \cite{zhang11a}.

The remainder of this Letter is organized as follows. First, we briefly review the CVM
using the $e^{+}$-potential scattering as an example. The reader is, however, referred
to the papers \cite{mitroy08d,zhang08c,zhang12c} for a full account. Second, we
numerically verify the CVM by calculating the phase shift and annihilation parameter of
$e^{+}$ scattering from an H atom, giving also a comparison to other methods. Then the
CVM is applied for studying the scattering of {\it o}-Ps from H$_2$ at low energy.

{\it Theory}.\ Phase shifts are expressed in radians and atomic units are used
throughout the following considerations, unless otherwise stated. Investigation of
elastic scattering of $e^{+}$ from a short-range spherically symmetric potential, which
\textquotedblleft represents\textquotedblright\ the short-range interaction between a
positron and spherical many-electron atom, essentially means solving the Schr\"{o}dinger
equation
\begin{equation}
\left( -\frac{\nabla^2}{2} + V_0(r) \right) \Psi_i(r) = E_i
\Psi_i(r) , \  \  E_i > 0. \label{se1ep}
\end{equation}
Assuming that the potential $V_0(r)$ is zero beyond a finite radius $R_0$, we may add a
confining potential $V_{\rm CP}(r)$ to the Hamiltonian in Eq.\ (\ref{se1ep}) in order to
convert a complicated problem of many-body continuum states into much easier problems of
many-body discrete bound states, one-dimension-potential bound states, and
one-dimension-potential scattering. The Schr\"{o}dinger equation of the confined
many-body system becomes
\begin{equation}
\left( -\frac{\nabla^2}{2} + V_0(r) + V_{\rm CP}(r) \right)
\Psi'_i(r) = E_i \Psi'_i(r). \label{se2}
\end{equation}
$V_{\rm CP}(r)$ is typically chosen in the form \cite{mitroy08d,zhang08c}
\begin{eqnarray}
V_{\rm CP}(r) &=& 0, \ \ \ \ r < R_0 ,\\
V_{\rm CP}(r) &=& G (r - R_0)^2, \ \ \ \ r \geq R_0,
\label{ACP}
\end{eqnarray}
where $G$ is a positive number. Confining potentials of this type are chosen to avoid
disturbing the $e^{+}$-potential interaction. Taking the discrete energies
$E_i$ and expectation values $\langle \Psi'_i(r) | V_{\rm CP} |\Psi'_i(r) \rangle $ as
reference, we tune the auxiliary one-dimensional potential $V_{\text{aux}}(r)$ by
solving the Schr\"{o}dinger equation
\begin{equation}
\left( -\frac{\nabla^2}{2} +  V_{\text{aux}}(r) + V_{\rm CP}(r)
\right) \Phi'_i(r)  =  E'_i \Phi'_i(r). \label{se4}
\end{equation}
Like $V_0(r)$, $V_{\text{aux}}(r)$ has to satisfy the boundary condition
$V_{\text{aux}}(r)=0$ for $r\geq R_0$. The purpose of tuning $V_{\text{aux}}(r)$ is to
achieve $E'_i=E_i$ and $\langle\Psi'_i(r)|V_{\rm CP}|\Psi'_i(r)\rangle=\langle\Phi'_i(r)|V_{\rm CP}|\Phi'_i(r)\rangle$.
To this aim, $V_{\text{aux}}(r)$ can be made flexible by inclusion of two or more
parameters to adjust its shape and strength. For the elastic scattering of $e^{+}$ from
an H atom, for example, $V_{\text{aux}}(r)$ is chosen in this work in the form
\begin{eqnarray}
V_{\text{aux}}(r) &&= V_{\lambda_i,\alpha_i,\xi_i,\beta_i}(r)  \label{cplxmp1} \\
&&= \lambda_i(1+1/r)\exp(-\alpha_i r) + \xi_ir^2 \exp(-\beta_i r^2), \nonumber
\end{eqnarray}
where $\lambda_i$, $\alpha_i$, $\xi_i$, and $\beta_i$ are the adjustable parameters.
Equality of the energies means that the phase shift is the same and equality of
$\langle V_{\rm CP} \rangle$ ensures that the normalization condition at the boundaries
is the same. Finally, the phase shift is obtained by solving the Schr\"{o}dinger
equation
\begin{equation}
\left( -\frac{\nabla^2}{2} +  V_{\text{aux}}(r) \right) \Phi_i(r) =
E_i \Phi_i(r). \label{se5a}
\end{equation}
The key point of the CVM is that the logarithmic derivatives of the wave functions
$\Psi_i(r)$, $\Psi'_i(r)$, $\Phi'_i(r)$, and $\Phi_i(r)$ are exactly the same for the
same energy $E_i$ at $R_0$, {\it i.\ e.},
\begin{eqnarray}
&& \Gamma_{\Psi_i}(R_0) \equiv\frac{1}{\Psi_i(R_0)}\frac{d\Psi_i}{dr} \bigg|_{R_0},\\
&& \Gamma_{\Psi_i}(R_0) =  \Gamma_{\Psi'_i}(R_0) =
\Gamma_{\Phi_i}(R_0) = \Gamma_{\Phi'_i}(R_0). \label{logderiv1}
\end{eqnarray}
In addition, the phase shift is a function of $\Gamma_{\Psi_i}(R_0)$, {\it i.\ e.},
$\delta_{0}(E_i) = f(\Gamma_{\Psi_i}(R_0))$. Therefore, the phase shift obtained from
solving Eq.~(\ref{se5a}) equals that of $e^{+}$-$V_0(r)$ scattering.

The calculation of the annihilation parameters $Z_{\rm eff}$ for $e^{+}$ scattering and
$^1\!Z_{\rm eff}$ for {\it o}-Ps scattering depends on the normalization of $\Psi'_i(r)$
to the scattering boundary condition. For $e^{+}$-H scattering, for example, the
procedure for determining $Z_{\rm eff}$ is as follows. First, the expectation value of
$\delta({\bf r}_{e^{-}}-{\bf r}_{e^{+}} )$ is computed with $\Psi'_i$,
\begin{eqnarray}
&&\langle \delta( { \mathbf r}_{e^{-}} - {\bf r}_{e^{+}} ) \rangle =  \nonumber \\
&& \langle \Psi^{'}_i({\bf r}_{e^{-}},{\bf r}_{e^{+}})|
\delta( {\bf r}_{e^{-}} - {\bf r}_{e^{+}} )| \Psi'_i({\bf r}_{e^{-}},{\bf r}_{e^{+}})\rangle.
\label{delta1}
\end{eqnarray}
Second, the ratio of $\Phi'_i(r)$ and the continuum radial wave function at $R_0$ is
computed. For $\text{S}$-wave scattering this is
$A_b={\Phi'_i(R_0)}/{(\sqrt{4\pi}\sin(kR_0+\delta_0))}$. Then $Z_{\rm eff}$ is defined
as
\begin{equation}
Z_{\rm eff}(k)=\frac{\langle\delta({\mathbf r}_{e^{-}}-{\bf r}_{e^{+}})\rangle}{A_b^2k^2}.
\end{equation}

{\it Scattering of} ${e}^+$ {\it from an H atom}.\ To demonstrate the accuracy of $A_b$
in the CVM, we calculate $\delta_{0}$, $A_b$, and $Z_{\rm eff}$ for the S-wave ${e}^+$-H
scattering at $k=0.2a_0^{-1}$, using two sets of basis functions: inner and outer. The
inner basis functions are chosen as ECG functions,
\begin{eqnarray}
\phi_k &=& \exp\left(-\frac{1}{2}\sum^{}_{ij}b_{k,ij}{\bf r}_i\cdot {\bf r}_j \right).
\end{eqnarray}
They are optimized using the stochastic variational method
\cite{kukulin77,varga95,suzuki98a,zhang08b}. The outer basis functions are expressed in
the form
\begin{eqnarray}
\Psi^{i}_{\rm outer} & = & \psi^{\rm H}({\bf r}_{e^-}) \exp{ \left( -\frac{1}{2} \alpha_i {\bf r}_{e^+}^2 \right)}, \\
    \psi^{\rm H}({\bf r}_{e^-}) & = & \sum_j d_j \exp\left(-\frac{\mu_j{{\bf r}_{e^-}^2}}{2} \right).
\label{out}
\end{eqnarray}
The wave function of the H ground state, $\psi^{\rm H}({\bf r}_{e^-})$, is written as
linear combination of 20 ECG functions with energy $E_{\text{H}}=-0.499\,999\,999\,43$
Hartree. Moreover, $\alpha_i $ is defined by the relation $\alpha_i = 18.6/1.45^{i-1}$
for $1 \leq i \leq 40$. To take into account the polarization effect of H,
$V_{\rm aux}(r)$ additionally includes the polarization potential
\begin{equation}
V_{\rm pol}(r) = -\frac{\alpha_d}{2r^4}\left( 1 - \exp(-r^6 / r_0^6) \right), \label{VH}
\end{equation}
with the static dipole polarizability $\alpha_d = 4.5$ a.u.\ and cut-off parameter
$r_0=2.16a_0$.

\begin{table}[t]
\begin{center}
\caption{Convergence of the results for S-wave ${e}^+$-H scattering at $k=0.2a_0^{-1}$
as function of the number $N$ of ECG functions. $k$: wave number; $\delta_{0}$: phase
shift; $Z_{\rm eff}$: annihilation parameter.}\label{hpswave}
\begin{tabularx}{0.48\textwidth}{lYYc}
\hline\hline
   $N$   &  $k$ ($a_0^{-1}$)  & $\delta_{0}$ (rad) & $Z_{\rm eff}$\\
\hline
 \multicolumn{1}{l}{$N_{\text{inner}}$}   & & & \\
200     & 0.20000185   &0.187536     & 5.482 \\
300     & 0.20000036  &0.187630    & 5.536 \\
400     & 0.20000011 &0.187646   & 5.545 \\
500     & 0.20000009  &0.187648   & 5.554 \\\hline
 \multicolumn{2}{l}{$N_{\text{inner}}+N_{\text{outer}}$}   & &      \\
240   &  0.20000072  & 0.187608   & 5.480 \\
340   &  0.20000012    & 0.187646   & 5.535 \\
440   & 0.20000002  & 0.187653    & 5.545\\
540   & 0.20000000  & 0.187654 & 5.553\\\hline
COP \cite{bhatia74b}  & 0.200  &0.1877    & 5.538  \\
{KV \cite{humberston72a,humberston97a} } & 0.200   & 0.1875    &   \\
{HNV \cite{gien99a} } & 0.200 & 0.1876  &  \\
{TM \cite{mitroy95b,ryzhikh00a}} & 0.200 &0.1868 &   5.5394  \\
\hline\hline
\end{tabularx}
\end{center}
\end{table}

Table~\ref{hpswave} addresses the convergence of our calculations for S-wave ${e}^+$-H
scattering and gives a comparison with other methods. We obtain
$k=\sqrt{2(E_3-E_{\text{H}})}$ from the third eigen-energy $E_3$ of the ${e}^+$-H system
confined in the potential $V_{\rm CP}(r_{e^+}) = G (r_{e^+}-R_0)^2$, where
$G = 2.732\,96\!\times\!10^{-5}$ and $R_0=21.0a_0$. For increasing size of the
inner basis, $G$ is tuned gradually so that $k$ approaches $0.2a_0^{-1}$. Then,
using $k$ and $\langle \Psi'_3 | V_{\rm CP} |\Psi'_3 \rangle$ as reference, we
determine the parameters of
$V_{\rm aux}(r)$ in Eq.\ (\ref{cplxmp1}). Keeping $\lambda_i=0.999\,50 $,
$\alpha_i=2.0a^{-1}_0 $, and $\beta_i=0.230 a^{-2}_0$ fixed for calculations including
the 40 outer basis functions, the requirement
$\langle\Psi'_i(r)|V_{\rm CP}|\Psi'_i(r)\rangle=\langle\Phi'_i(r)|V_{\rm CP}|\Phi'_i(r)\rangle$
can be satisfied by tuning only $\xi_i$. The operator $\delta( {\bf r}_{e^{-}}-{\bf r}_{e^{+}})$
does not commute with the Hamiltonian so that there are no common eigen-states.
During the optimization, many sets of nonlinear parameters may give the same energy but
they generate different expectation values
$\delta_{ep}=\langle\delta({\bf r}_{e^{-}}-{\bf r}_{e^{+}})\rangle$. Therefore, the
energy is variationally minimized, while $\delta_{ep}$ is variationally maximized.

The convergence of $k$ is accelerated by augmenting the outer basis. As a
consequence, the convergence of $\delta_0$, which is related to $k$, is also
accelerated. However, this procedure makes $\delta_{ep}$ slightly smaller than
calculated with only the inner basis. Both  $k$ and $\delta_{0}$ show very good
convergence. We obtain $\delta_{0} = 0.18765$ rad. This result agrees well with the
extrapolated value (0.1877 rad) of the correlated optical potential (COP) calculation
by Bhatia {\it et al.} \cite{bhatia71}, with the value (0.1875 rad) of the Kohn
variational (KV) calculation by Humberston {\it et al.} \cite{humberston97a}, and with
the value (0.1876 rad) of the Harris--Nesbet variational (HNV) calculation by
Gien~\cite{gien99a}. On the other hand, it is 4.6\textperthousand\ larger than the
value (0.1868 rad) of the 21-state close coupling approach \cite{mitroy95b}. As the
COP result (5.538 \cite{bhatia74b}) and T-matrix (TM) result (5.5394
\cite{mitroy95b,ryzhikh00a}) are close to the value calculated with 300 ECG function
(5.536), the CVM final result of $Z_{\rm eff}=5.553$ is more accurate than the results
of COP and TM methods. It turns out that $Z_{\rm eff}$ increases monotonically with the
number of ECG functions but converges slowly. We estimate that the exact value of
$Z_{\rm eff}$ falls within the range from 5.554 to 5.559. We note that calculation with
only a large inner basis has the capacity to generate accurate values for
$\delta_{0}$ and $Z_{\rm eff}$.

{\it Scattering of o-Ps from $H_2$}.\ We employ the fixed nucleus approximation
with an internuclear distance of $R_{\text{H}_2}=1.45a_0$, which is almost the
equilibrium distance $1.448a_0$. Moreover, $E_{\text{H}_2}=-1.174\,057\,038$ Hartree
as calculated by Rychlewski {\it et al.} \cite{rychlewski94a} with 300 ECG functions is
adopted for the ground state energy of $\text{H}_2$. The Hamiltonian for the
{\it o}-Ps-H$_2$ scattering is
\begin{eqnarray}
H &=& -\sum_{i=1}^4 \frac{\nabla_i^{2}}{2}
  + \sum_{j>i=1}^4\frac{q_i\,q_j}{|{\bf r}_j-{\bf r}_i|} \nonumber \\
  &+&  \sum_{i=1}^4 \left\{\frac{q_i}{|{\bf r}_i-{\bf R}/2|}
  + \frac{q_i}{|{\bf r}_i+{\bf R}/2|} \right\}, \label{scatHam}
\end{eqnarray}
where ${\bf r}_i$ is the coordinate of the $i$-th particle ($e^{\pm}$) relative to the
midpoint of the $\text{H}_2$ molecular axis and $q_i$ is its charge. The vectors
$\pm {\bf R}/2$ represent the displacements of the two protons from the midpoint. The
basis for the interaction region has the form
\begin{eqnarray}
\phi_k &=& {\hat P} \ \exp \left( -\frac{1}{2}  \sum^{4}_{i=1}b_{k,i}|{\bf r}_i - {\bf S}_{k,i}|^2 \right) \nonumber \\
    & \times & \exp \left( -\frac{1}{2} \sum^{3}_{i=1} \sum^{4}_{j=i+1} a_{k,ij}|{\bf r}_i - {\bf r}_{j}|^2  \right).
\end{eqnarray}
The vector ${\bf S}_{k,i}$ displaces the center of the ECG function for the $i$-th
particle to a point on the internuclear axis. The operator ${\hat P}$ ensures that the
basis has $\Sigma_g$ symmetry. The confining potential is added in the center-of-mass
coordinate ${\rhobo}_i =({\bf r}_{e^{+}}+{\bf r}_i)/2$ so that the potential acting on
the center-of-mass of $e^{+}$ and the $i$-th $e^{-}$ of the target is not reasonable.
However, this effect declines for increasing $R_0$~\cite{zhang12c}.

Following previous experience with the S-wave elastic scattering of Ps from
an H atom \cite{zhang12c}, $R_0=24a_0$ is used for the {\it o}-$\text{Ps-H}_2$
scattering. As {\it o}-Ps experiences during the scattering a van der Waals potential,
we choose the auxiliary potential as
\begin{equation}
V_{ \lambda_i,\alpha_i}(\rho) = \lambda_i  \exp(-\alpha_i\rho )  -
\frac{C_6}{\rho^6}\left( 1 - \exp(-\rho^6 / \rho_0^6) \right), \label{VH2}
\end{equation}
with cut-off parameter $\rho_0 = 6.0 a_0$ and dispersion coefficient $C_6=49.3$
a.u.~\cite{mitroy03e}. Only the inner basis is used, because the outer basis is too
complicated in this case. Similar to the case of $e^+$-H scattering, we expect that
accurate scattering parameters can be obtained with a large inner basis. In the
following text, we use a superscript $T$ to indicate the triplet spin character of the
pick-off annihilation. Due to the complexity of the multi-center scattering system, variational
optimization of the energy and $\delta^{T}_{ep}$ together is very time consuming. Hence,
only the energy is optimized by adjusting the nonlinear parameters of the ECG functions.

\begin{table}[ht]
\begin{center}
\caption{Convergence of the results for $\Sigma_g$ {\it o}-Ps-H$_2$ scattering at
$k=0.1 a^{-1}_0$ as function of the number $N$ of ECG functions. $k$: wave number;
$\delta^{T}_{ep}=\langle \delta^{T}( { \mathbf r}_{e^{-}} - {\bf r}_{e^{+}} ) \rangle$;
$\delta_{0}$: phase shift; $^1\!Z_{\rm eff}$: pick-off annihilation parameter.}\label{h2psk01}
\begin{tabularx}{0.48\textwidth}{lYYYc}
\hline\hline
   $N$   &  $k (a_0^{-1})$  & $\delta^{T}_{ep}$  & $\delta_{0}$ (rad) & $^1\!Z_{\rm eff}$    \\
\hline
2400  & 0.100061 &  $8.4043\!\times\!10^{-5}$ & $-0.1876$    & 0.1637 \\
2800  & 0.100018 &  $8.4635\!\times\!10^{-5}$ & $-0.1863$   & 0.1659 \\
3200  & 0.100006 &  $8.4687\!\times\!10^{-5}$ & $-0.1859$    & 0.1662 \\
3600  & 0.100002 &  $8.4873\!\times\!10^{-5}$ & $-0.1857$   & 0.1668 \\\hline\hline
\end{tabularx}
\end{center}
\end{table}

Table~\ref{h2psk01} addresses the convergence of the results for $\Sigma_g$
{\it o}-Ps-H$_2$ scattering at $k=0.1a_0$ when the number of ECG functions increases.
We have $k=2\sqrt{(E_1-E_{\text{Ps}}-E_{\text{H}_2})}$, where $E_1$ is generated with
the confining potential parameter $G=1.7666\!\times\!10^{-4}$ (obtained from the
optimization of the nonlinear parameters) and $E_{\text{Ps}}=-0.25$ Hartree is the exact
energy of the Ps ground state. In Eq.\ (\ref{VH}), $\lambda_i $ and $\alpha_i$ have
to be tuned together for each basis to satisfy the requirements to $k$ and
$\langle\Phi'_1(r)|V_{\rm CP}|\Phi'_1(r)\rangle$. For a basis with 3600 ECG functions,
for example, we obtain $\lambda_i\simeq-0.382\,742$ and $\alpha_i\simeq0.553$. Both $k$
and $\delta_{0}$ show good convergence for an increasing number of ECG functions, in
contrast to $\delta^{T}_{ep}$ and $^1\!Z_{\rm eff}$ (though they vary monotonically).
%%%
\begin{table}[ht]
\begin{center}
\caption{ \label{h2psks}
Confining potential parameter $G$, $\delta^{T}_{ep}=\langle \delta^{T}({\mathbf r}_{e^{-}} - {\bf r}_{e^{+}} ) \rangle$,
phase shift $\delta_{0}$, and pick-off annihilation parameter $^1\!Z_{\rm eff}$ for
$\Sigma_g$ {\it o}-Ps-H$_2$ scattering at different $k$. Experimental values of
$^1\!Z_{\rm eff}$ are listed for comparison. Numbers in parentheses give the
uncertainty in the last digit.}
\begin{tabularx}{0.48\textwidth}{YYYYc}
\hline\hline
\multicolumn{1}{c}{$k$ ($a_0^{-1}$)} & \multicolumn{1}{c}{$G$} & \multicolumn{1}{c}{$\delta^{T}_{ep}$}  &\multicolumn{1}{c}{$\delta_{0}$ (rad)}
&\multicolumn{1}{c}{$^1\!Z_{\rm eff}$}      \\
\hline
\multicolumn{1}{l}{0.06098} & $2.15\!\times\!10^{-6}$  & $1.7904\!\times\!10^{-6}$ & $-0.1215$    & 0.1737 \\
\multicolumn{1}{l}{0.08280} & $2.27\!\times\!10^{-6}$  & $4.7099\!\times\!10^{-5}$ & $-0.1547$   & 0.1687 \\
\multicolumn{1}{l}{0.10000} & $1.7666\!\times\!10^{-4}$& $8.4873\!\times\!10^{-6}$ &  $-0.1857$  & 0.1668 \\\hline
\multicolumn{1}{l}{0.0} & \multicolumn{3}{l}{Effective-range theory} &     $0.1858$     \\\hline
\multicolumn{4}{l}{Exp.\ at 77.4 K \cite{mcnutt79a}} &   $0.197(3)$    \\
\multicolumn{4}{l}{Exp.\ at 250 K \cite{mcnutt79a}}  &   $0.195(5)$    \\
\multicolumn{4}{l}{Exp.\ at 293 K \cite{mcnutt79a}}  &   $0.193(5)$    \\
\multicolumn{3}{l}{Exp.\ at 293 K \cite{wright83a}}  &   &  $0.186(1)$     \\
\hline\hline
\end{tabularx}
\end{center}
\end{table}

Table~\ref{h2psks} presents results of our CVM calculations for three values of $k$. We
focus our attention on scattering with $k \leq 0.1 a_0$ for two reasons. First, the most
reliable experimental information comes from annihilation experiments of thermal
{\it o}-Ps. Second, the collision can be treated as S-wave scattering and, thus, the
molecular aspects of the asymptotic wave function can be neglected. By fitting
$\delta_0$ from Table~\ref{h2psks} to the effective-range theory \cite{drake06a},
\begin{eqnarray}
k\text{cot}(\delta_k) = -\frac{1}{A_s}+\frac{1}{2}r_0k^2 -\frac{4\pi C_6}{15 A^2_s}k^3
                        -\frac{16C_6}{15 A_s}k^4\ln(k), \nonumber \\ \label{eft}
\end{eqnarray}
the scattering length $A_s=2.06a_0$ and effective range $r_0=9.715a_0$ are obtained.
The value of $A_s=(2.1\pm0.2)a_0$ estimated from the average experimental 
momentum-transfer cross section for Ps energy below 0.3 eV \cite{nagashima03a}
agrees well with this result. The value of the pseudopotential method is much smaller
($0.64 a_0$) \cite{wilde15a}. In addition, our result for the S-wave cross section at
$k=0.1 a_0$ ($13.59\pi a_0^2$) is much larger than the value of the three-Ps-state
coupled-channel method ($3.79\pi a_0^2$) \cite{biswas00b}. This means that both these
methods seriously underestimate the near-zero-energy cross section.

Using the effective-range theory expansion \cite{mitroy02e},
\begin{eqnarray}
{^1\!Z_{\rm eff}(k)} = {^1\!}Z^{(0)}_{\rm eff} + {^1\!}Z^{(1)}_{\rm eff}k^2 + {^1\!}Z^{(2)}_{\rm eff}k^3,
 \end{eqnarray}
fitting leads to ${^1\!}Z^{(0)}_{\rm eff}=0.186$. Experimental values of $0.197(3)$
\cite{mcnutt79a}, $0.195(5)$ \cite{mcnutt79a}, $0.193(5)$ \cite{mcnutt79a}, and
$0.186(1)$~\cite{wright83a} from weighted least-squares fits of observed decay rates at
low $\text{H}_2$ gas densities and temperatures of 77.4 K, 250 K, 293 K, and 293 K,
respectively, indicate that the low-density ${^1\!}Z^{(0)}_{\rm eff}$ is
independent of the temperature (at the level of accuracy of the experimental data). The
fit of Ref.~\cite{mcnutt79a} was constrained to a vacuum annihilation rate of
$\Gamma_{\text{vac}}=7.24$ $\mu\text{sec}^{-1}$, which is about $2.8\%$ larger than the
experimental value of 7.0401(7) $\mu\text{sec}^{-1}$~\cite{kataoka09a}. Using no such
constraint, a better value of $\Gamma_{\text{vac}}=6.95(8)$ $\mu\text{sec}^{-1}$ was
determined in Ref.\ \cite{wright83a}. We obtain perfect agreement with the
experimental value of ${^1\!}Z^{(0)}_{\rm eff}=0.186(1)$ from Ref.\ \cite{wright83a}.

{\it Summary}.\ The CVM is a powerful method that fully utilizes the advantages of
studying bound states of atoms and molecules to determine phase shifts and normalization
constants of asymptotic wave functions for collisions. The accuracy of the CVM
normalization constant has been verified for $e^+$-H scattering by comparison with other
methods. The CVM result of $Z_{\rm eff}=5.553$ for S-wave $e^+$-H scattering at
$k=0.2a_0^{-1}$ is the first significant improvement in accuracy since the COP value of
$Z_{\rm eff}=5.538$ was reported in 1974 \cite{bhatia74b}. For {\it o}-Ps-H$_2$
scattering, we have reported accurate values of $\delta_{0}$ and $^1\!Z_\text{eff}$ for
three different incident momenta. The CVM results for ${^1\!}Z^{(0)}_{\rm eff}$ and
$A_s$, extracted by means of the effective-range theory, show excellent agreement with precise experimental
data \cite{wright83a,nagashima03a}. As this problem was intractable for a long time, we
believe that the present study will inspire new theoretical and experimental efforts on
the low-energy {\it o}-Ps scattering from few-body targets.

{\it Acknowledgement}.\ J.Y.Z. is grateful to X. Gao and colleagues for their
hospitality during his visit to Beijing Computational Science Research Center. Z.C.Y.
was supported by the NSERC of Canada and in part by the Chinese Academy of Sciences
CAS/SAFEA International Partnership Program for Creative Research Teams.
The research reported in this publication was supported by funding from King Abdullah
University of Science and Technology (KAUST). The authors are thankful for the
computational resources provided by the ACEnet of Canada.
\end{document}